\pdfoutput=1
\documentclass[aps,prb,reprint,superscriptaddress,longbibliography]{revtex4-1}

\usepackage{graphicx}
\usepackage{amsmath}
\usepackage{amssymb}
\usepackage{amsfonts}
\usepackage{dcolumn}
\usepackage{dsfont}
\usepackage{latexsym}
\usepackage{rotating}
\usepackage{color}
\usepackage{latexsym}
\usepackage{bbm}
\usepackage{subfigure}
\usepackage{float}
\usepackage{epsfig}
\usepackage{psfrag}
\usepackage{natbib, hyperref}
\usepackage{bm}
\usepackage{amsthm}
\usepackage{eucal}
\usepackage{mathrsfs}
\usepackage{url}
\usepackage{braket}
\usepackage{ulem}
\usepackage{calligra}
\usepackage[T1]{fontenc}
\usepackage[utf8]{inputenc}
\usepackage{newunicodechar}
\usepackage{marvosym}
\usepackage[absolute]{textpos}
\usepackage[cal=cm]{mathalfa}
\usepackage[resetlabels]{multibib}  
\newcites{Supp}{Supplementary~References} 
\usepackage{color}

\usepackage{hyperref}
\hypersetup{
colorlinks=true,final=true,
        linkcolor=blue,
        citecolor=blue,
        filecolor=blue,
        urlcolor=blue,
}
\usepackage{xcolor}
\colorlet{myPurple}{blue!40!red}
\definecolor{myOrange}{rgb}{1,0.5,0}

\begin{document}

\title{Signatures of a liquid-crystal transition in spin-wave excitations of skyrmions}

\author{Narayan Mohanta$^*$}
\affiliation{Materials Science and Technology Division, Oak Ridge National Laboratory, Oak Ridge, TN 37831, USA}
\author{Andrew D. Christianson}
\affiliation{Materials Science and Technology Division, Oak Ridge National Laboratory, Oak Ridge, TN 37831, USA}
\author{Satoshi Okamoto}
\affiliation{Materials Science and Technology Division, Oak Ridge National Laboratory, Oak Ridge, TN 37831, USA}
\author{Elbio Dagotto}
\affiliation{Materials Science and Technology Division, Oak Ridge National Laboratory, Oak Ridge, TN 37831, USA}
\affiliation{Department of Physics and Astronomy, The University of Tennessee, Knoxville, TN 37996, USA\\
\rm $^*$Email: mohantan@ornl.gov}

\begin{abstract}
\noindent {\textbf {Abstract:}} 
Understanding the spin-wave excitations of chiral magnetic order, such as the skyrmion crystal (SkX), is of fundamental interest to confirm such exotic magnetic order. The SkX is realized by competing Dzyaloshinskii-Moriya and ferromagnetic-exchange interactions with a magnetic field or anisotropy. Here we compute the dynamical spin structure factor, using Monte Carlo and spin dynamics simulations, extracting the spin-wave spectrum in the SkX, in the vicinity of the paramagnet to SkX transition. Inside the SkX, we find six spin-wave modes, which are supplemented by another mode originating from the ferromagnetic background. Above the critical temperature $T_s$ for the skyrmion crystallization, we find a diffusive regime, reminiscent of the liquid-to-crystal transition, revealing that topological spin texture of skyrmionic character starts to develop above $T_s$ as the precursor of the SkX. We discuss the opportunities for the detection of the spin waves of the SkX using inelastic-neutron-scattering experiments in manganite-iridate heterostructures.
\end{abstract}
       
\maketitle

\noindent {\bf Introduction\\}
Chiral magnetic order occurring in several condensed-matter systems has added a new dimension to the realization of novel functionalities and exotic fundamental physics~\cite{Barron_NMat2008,Fert_NRevMat2017}. Magnetic skyrmions, the vortex-like spin swirls, belong to the class of chiral magnetic textures having a quantized topological charge~\cite{Nagaosa_NNat2013}. In the presence of competing spin exchange interactions, the skyrmions often arrange themselves naturally in a triangular array to form a skyrmion crystal (SkX) in real lattices~\cite{Ezawa_PRB2011,Buhrandt_PRB2013,Rohart_PRB2013,Motome_PRL2017}. The existence of the SkX phase has been established in a surge of recent experiments by Lorentz transmission electron microscopy, spin-resolved scanning tunneling microscopy, and elastic small-angle neutron scattering (SANS)~\cite{Rossler_Nature2006,Muhlbauer_Science2009,Yu_Nature2010,Yu_NMat2011,Heinze_NPhys2011,Yu_NComm2014,Park_NNat2014,Woo_NMater2016,Herve_NComm2018,Muhlbauer_RMP2019}. A magnetic skyrmion generates an emergent electro-magnetic field~\cite{Schulz_Nphys2012,Nagaosa_PhilTrans2012}, giving rise to topological Hall effect in metals. This effect also serves as an indirect yet effective probe for the chiral magnetic order in the SkX phase~\cite{Neubauer_PRL2009,Kanazawa_PRL2011,Taguchi_Science2001,Huang_PRL2012,Rana_NJP2016,Wang_NMat2018,Hamamoto_PRB2015,Yi_PRB2009,Nakamura_JPSJ2018,Swekis_PRM2019,Vir_PRB2019,NM_PRB2020_PTHE}. 

The SkX phase exhibits spin-wave excitations which are Goldstone modes associated with the spontaneously-broken translational symmetry of the chiral spin arrangement. The dispersion of these spin-wave modes is determined by the properties of the SkX. Three SkX spin-wave modes -- viz. clockwise, counterclockwise circulation and breathing modes -- have been predicted in previous theoretical analysis~\cite{Mochizuki_PRL2012,Petrova_PRB2011,Garst_JAPD2017} and subsequently reported in experiments using broadband microwave spectroscopy~\cite{Schwarze_NMat2015,Onose_PRL2012,Ehlers_PRB2016}. Some of the spin-wave modes exhibit non-trivial topological properties, revealing in the form of chiral edge states in the SkX~\cite{Rossier_NJP2016}. The transition to the SkX phase takes place within a range of external magnetic field strengths, when the temperature is slowly lowered down~\cite{NM_PRB2019}.

An important puzzle which remained unresolved for years is the role of spin fluctuations in the transition to the SkX phase. This is particularly interesting since specific heat measurements report a first-order phase transition~\cite{Bauer_PRL2013} to the helimagnetic phase at zero magnetic field while neutron-scattering experiments reveal a Landau soft-mode mechanism of weak crystallization to the SkX phase at a finite field~\cite{Wilhelm_PRL2011,Samatham_PhysStatusSol2013,Kindervater_PRX2019}. These experiments suggest that the emergence of the SkX phase occurs via a precursor regime in which non-trivial topological spin textures of skyrmionic character are developed within a paramagnetic background with abundance of fluctuations, similar to the precursor phenomenon proposed in the context of liquid-crystal transitions~\cite{Brazovskii_JETP1987}. At low finite temperatures, the positional correlations of the skyrmions decay with distance as power laws while the orientational correlations remain finite~\cite{Nishikawa_PRB2019}. With increasing temperature, a transition takes place from the solid to a liquid phase of skyrmions with asymptotically vanishing long-range correlations.

Here, we theoretically investigate the spin-wave excitations of a two-dimensional SkX phase and its evolution at higher temperatures, in the vicinity of the paramagnetic-to-SkX phase transition, by computing the dynamical spin structure factor $S(\mathbf{q},\omega)$, expecting its future direct detection in inelastic SANS experiments. We use a spin Hamiltonian to first find out the stable spin configurations at low temperatures using the Metropolis Monte Carlo annealing method and then numerically solve the Landau-Lifshitz equation of motion to capture the spin dynamics. The computed $S(\mathbf{q},\omega)$ reveals an exotic array of six gapless spin-wave modes in the SkX phase, stabilized within a range of the external perpendicular magnetic fields, accompanied by another gapless mode originating from the ferromagnetic background. In the diffusive regime for the skyrmion crystallization, $S(\mathbf{q},\omega=0)$ profile reveals a clear transition from a ring-shaped profile to a six-peak profile, characteristic of the SkX phase, confirming the previous experimental findings that the transition to the SkX phase upon cooling slowly from a higher temperature occurs via a precursor regime. A gradual decrease in the $S(\mathbf{q=0},\omega)$ intensity with temperature, at the SkX spin-wave modes, also indicates a phase transition similar to the liquid-crystal transition.

The spin spiral (SS) phase, which is the natural solution at zero magnetic field, exhibits two gapless spin-wave modes from the spiral order and a soft ferromagnetic mode that appears at finite energies and goes away with increasing the DMI strength. The field-polarized ferromagnetic (FM) phase exhibits a spin-wave mode which appears with an excitation gap due to the Larmor precession in a magnetic field. 

A particularly interesting platform that hosts a two-dimensional SkX is oxide interfaces, such as the interface between the ferromagnetic metal La$_{1-x}$Sr$_{x}$MnO$_3$ ($0.2\! \lessapprox  \!x\! \lessapprox \!0.5$) and the non-magnetic semimetal SrIrO$_3$. The Dzyaloshinskii-Moriya interaction (DMI), which arises at this interface due to the strong spin-orbit coupling in SrIrO$_3$ and the broken inversion symmetry, stabilizes a N\'eel-type SkX~\cite{NM_PRB2019}. We discuss the feasibility of the experimental detection of our results in this manganite/iridate interfaces based on the state-of-the-art SANS spectrometers and the developments required for a successful future confirmation of our predictions.\\

\begin{figure}[t]
\begin{center}
\epsfig{file=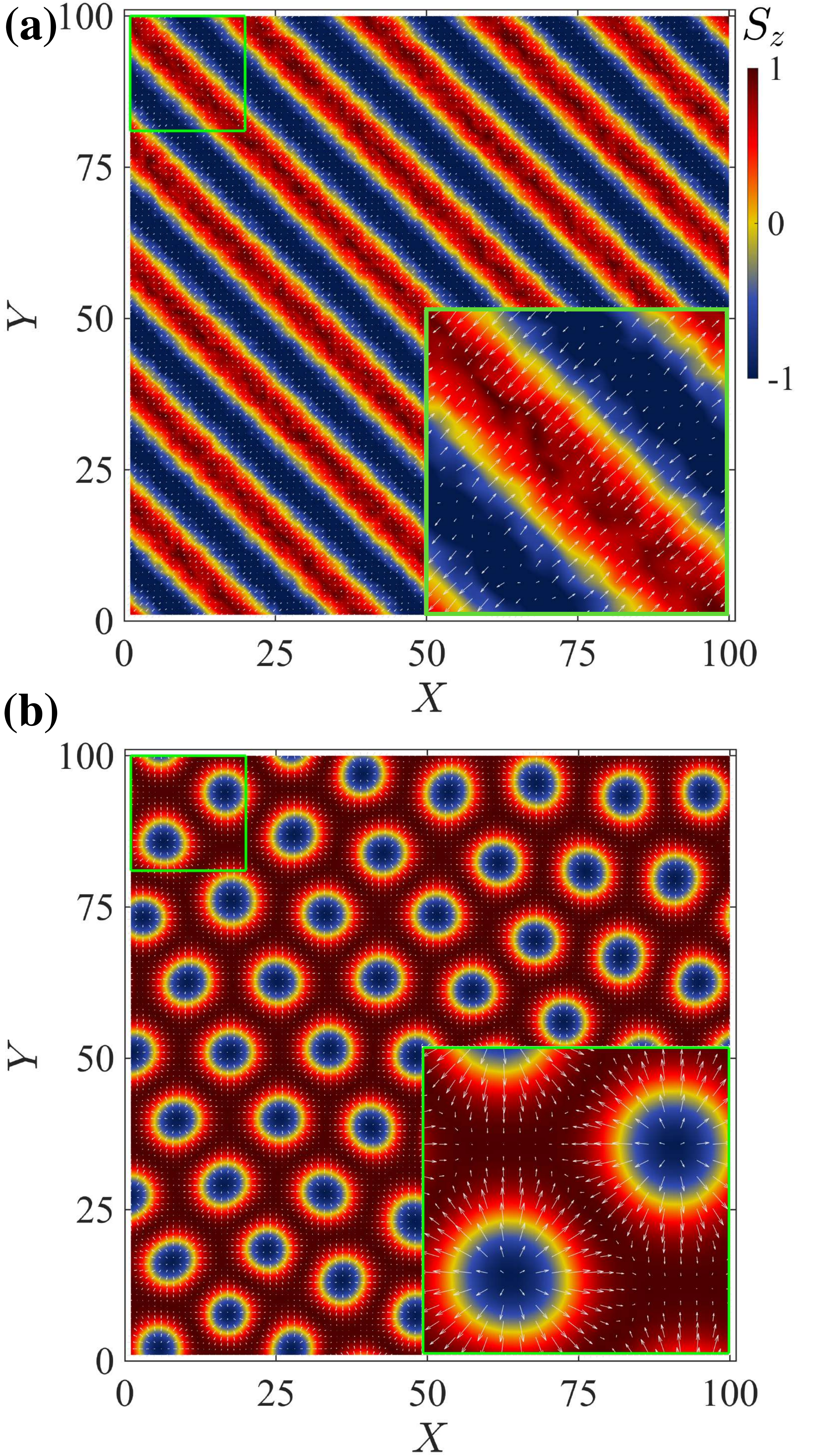,trim=0.0in 0.0in 0.0in 0.0in,clip=false, width=80mm}
\caption{{\bf Spin textures in the spin spiral (SS) and the skyrmion crystal (SkX) phases.} The spin textures were spontaneously generated in the Monte Carlo annealing process in a 100$\times$100 square lattice with periodic boundaries: (a) SS texture realized at zero magnetic field ($H_{z}\!=\!0$), (b) nearly-triangular crystal of skyrmions realized at a magnetic field $H_{z}\!=\!0.18J$ ($\approx 3.74$~T). Parameters used are ferromagnetic exchange interaction strength $J\!=\!1$, spin amplitude $S\!=\!3/2$, Dzyaloshinkii-Moriya interaction strength $D\!=\! 0.5J$, easy-plane anisotropy amplitude $A\!=\!0.01J$ and temperature $T\!=\!0.001J$ ($\approx 14$~mK). The white arrows and the colorbar respectively show the in-plane ($x$ and $y$) components and the perpendicular ($z$) component of the spins. The insets show an expanded view of the spin configuration inside the region denoted by the green square on the top-left corner.}
\label{Spin_Config}
\vspace{-2em}
\end{center}
\end{figure}
\noindent {\bf Results\\}
\label{sectionII}
\noindent {\bf Model Hamiltonian.} We consider a model Hamiltonian, relevant to two-dimensional material systems such as the La$_{1-x}$Sr$_{x}$MnO$_3$/SrIrO$_3$, Pt/Co and Fe/Ir interfaces. We consider the manganite/iridate interface as our example system and choose parameters that are relevant to this interface. The manganite is considered to be doped suitably to be in the ferromagnetic phase in which the localized spins can be described effectively by a Heisenberg exchange interaction. The influence of the strong spin-orbit coupling of the iridate, in the absence of explicit structural inversion symmetry, leads to the DMI at the interface. The total Hamiltonian, in the presence of an external magnetic field, is given by
\begin{align}
\mathcal{H}\!=\!&-J\sum_{\langle ij \rangle} \mathbf{S}_i \cdot \mathbf{S}_j -D \sum_{\langle ij \rangle}(\hat{z} \times \hat{r}_{ij}) \cdot ( \mathbf{S}_i \times \mathbf{S}_j  ) \nonumber \\
&-H_z\sum_{i}S_{zi}-A\sum_{i}|S_{zi}|^2,
\label{Hamiltonian}
\end{align}

\begin{figure*}[t]
\begin{center}
\epsfig{file=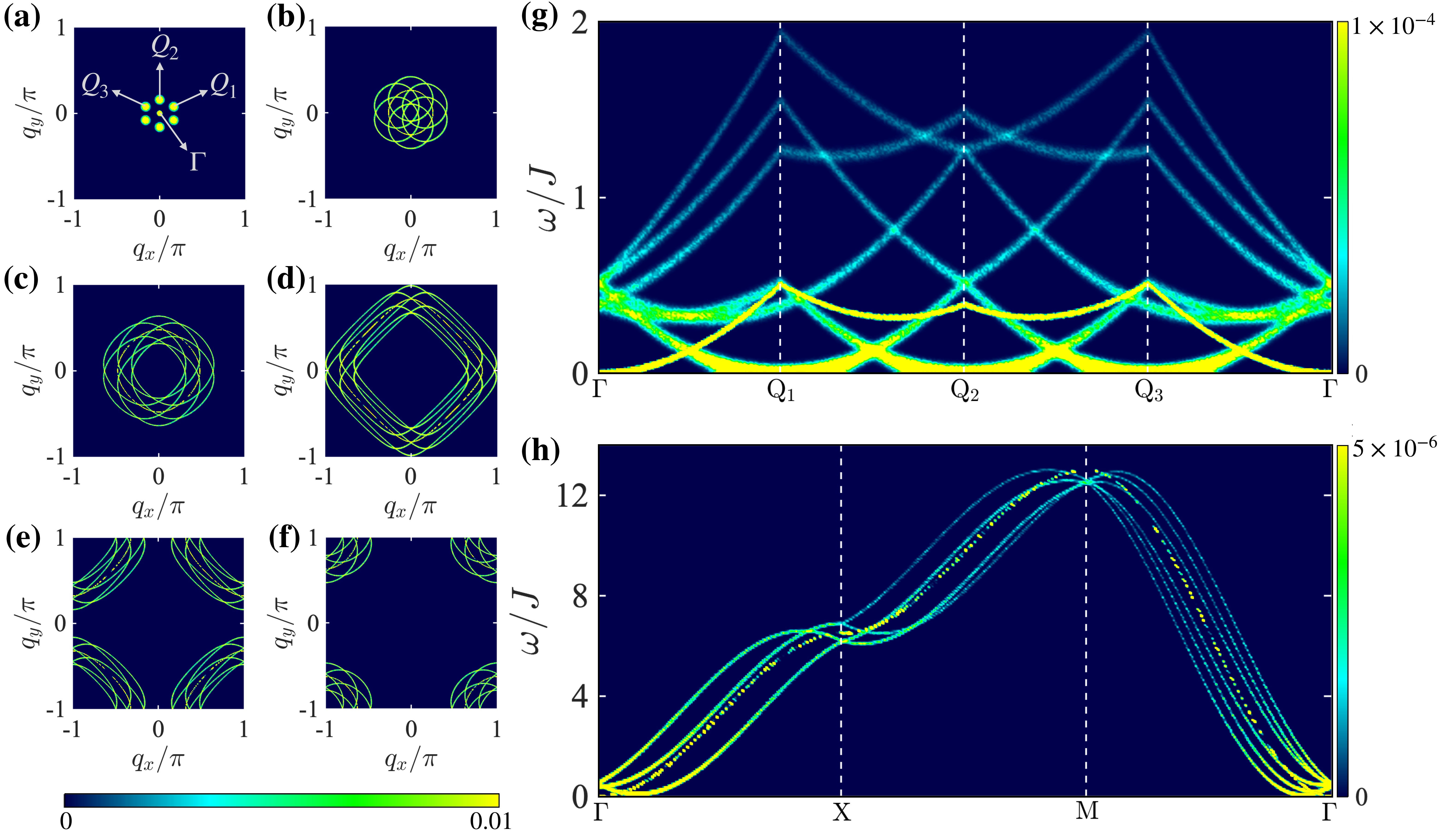,trim=0.0in 0.0in 0.0in 0.0in,clip=false, width=180mm}
\caption{{\bf Spin waves in the skyrmion crystal (SkX) phase.} (a)-(f) Constant-energy contours of the dynamical spin structure factor $S(\mathbf{q},\omega)$ calculated in the momentum plane $\mathbf{q}=(q_x,q_y)$ in the SkX phase, realized at a magnetic field $H_{z}\!=\!0.18J$ ($\approx 3.74$~T), at energies (a) $\omega \!=\!0$, (b) $\omega \!=\!J$, (c) $\omega \!=\!3J$, (d) $\omega \!=\!6J$, (e) $\omega \!=\!8J$, and (f) $\omega \!=\!11J$. Plot (a) shows the three characteristic momenta $\mathbf{Q}_1$, $\mathbf{Q}_2$ and $\mathbf{Q}_3$ of the SkX. (g)-(h)  Plots of $S(\mathbf{q},\omega)$ along the momentum paths (g) $\Gamma - \mathbf{Q}_1 - \mathbf{Q}_2 - \mathbf{Q}_3 - \Gamma$ and (h) $\Gamma - \rm X - \rm M - \Gamma$. The spin-wave mode originating from the $\Gamma$ (0,0) point is due to the ferromagnetic background in the SkX. The rest six branches arise from the triangular SkX spin configuration. }
\label{SW_SkX}
\vspace{-2em}
\end{center}
\end{figure*}
where $J$ is the nearest-neighbor ferromagnetic exchange parameter, $D$ is the DMI strength, $H_z$ is the magnetic field applied perpendicular to the interfacial plane, and $A$ is the easy-plane magnetic anisotropy originating from the combined interfacial strain and Rashba spin-orbit coupling~\cite{Randeria_PRX2014}. We used $A\!=\!0.01J$ throughout this paper. $\mathbf{S}_i$ is the localized spin of amplitude $S\!=\!3/2$ on the Mn $t_{2g}$ orbitals at site $i$ on the manganite side of the interface. By comparing the critical temperature for the zero-field FM phase (at $D\!=\!0$) extracted from our Hamiltonian with the experimental findings~\cite{HoNyung_NComm2016}, it was found that the Heisenberg exchange energy constant at the manganite/iridate films varies within the range $1.2~{\rm meV} \lessapprox J  \lessapprox 3.9~{\rm meV}$, depending upon the thickness of the film~\cite{NM_PRB2019}. In the description below, we present energies in units of $J$ and reveal magnetic field and temperature scales, at selected places, using $J=1.2~{\rm meV}$. The dipole-dipole interaction usually has an important contribution in ferromagnetic materials. However, recent experiments suggest that the short-range exchange interaction dominates over the long-range dipole-dipole interaction in Sr-doped manganite compounds~\cite{Tiwari_JPCM2020}. While the long-range dipole-dipole interaction may be sizeable and help in stabilizing the magnetic phases in bulk manganite systems, here we focus on the short-range exchange interactions which are more relevant to the two-dimensional geometry.\\

\noindent {\bf Low-temperature spin configurations.} We present the spin-wave excitations, obtained using a $100\! \times \! 100$ lattice at different magnetic fields, first at a low temperature $T\!=\!0.001J$ and then discuss the high-temperature diffusive regime in which the liquid-crystal transition takes place. The SS phase, which is stabilized at zero magnetic field, evolves into a triangular SkX within a range of magnetic fields and finally becomes a fully-polarized FM phase at higher fields. A detailed analysis of the phase diagram, spanned by the magnetic field and the temperature, was done in our previous work~\cite{NM_PRB2019}. In Fig.~\ref{Spin_Config}(a) and (b), we show the  spin configurations of the SS phase and the SkX phase, obtained in a typical annealing session of the Monte Carlo simulations, respectively at zero field ($H_z\!=\!0$) and at a finite field ($H_z\!=\!0.18J$). Since the DM vectors lie predominantly at the interface plane, the SS and the SkX textures are N\'eel-type in nature, unlike in three-dimensional chiral magnets such as MnSi in which out-of-plane DM vectors lead to Bloch-type textures. A change in the sign of $D$ reverses the sense of rotation of the spins in going from the skyrmion center towards the radially outward direction. At the considered DMI strength $D\!=\!0.5J$, which is close to the value realizable at oxide interfaces~\cite{Matsuno_SciAdv2016,Wang_CommunPhys2018,Buttner2018}, the period of the SS and the skyrmion diameter are nearly 12 lattice spacings. At smaller values of $D$, these length scales increase almost exponentially, $D\!=\!0$ ($H_z\!=\!0$) being the rotationally-invariant Heisenberg FM phase. \\

\begin{figure*}[t]
\begin{center}
\epsfig{file=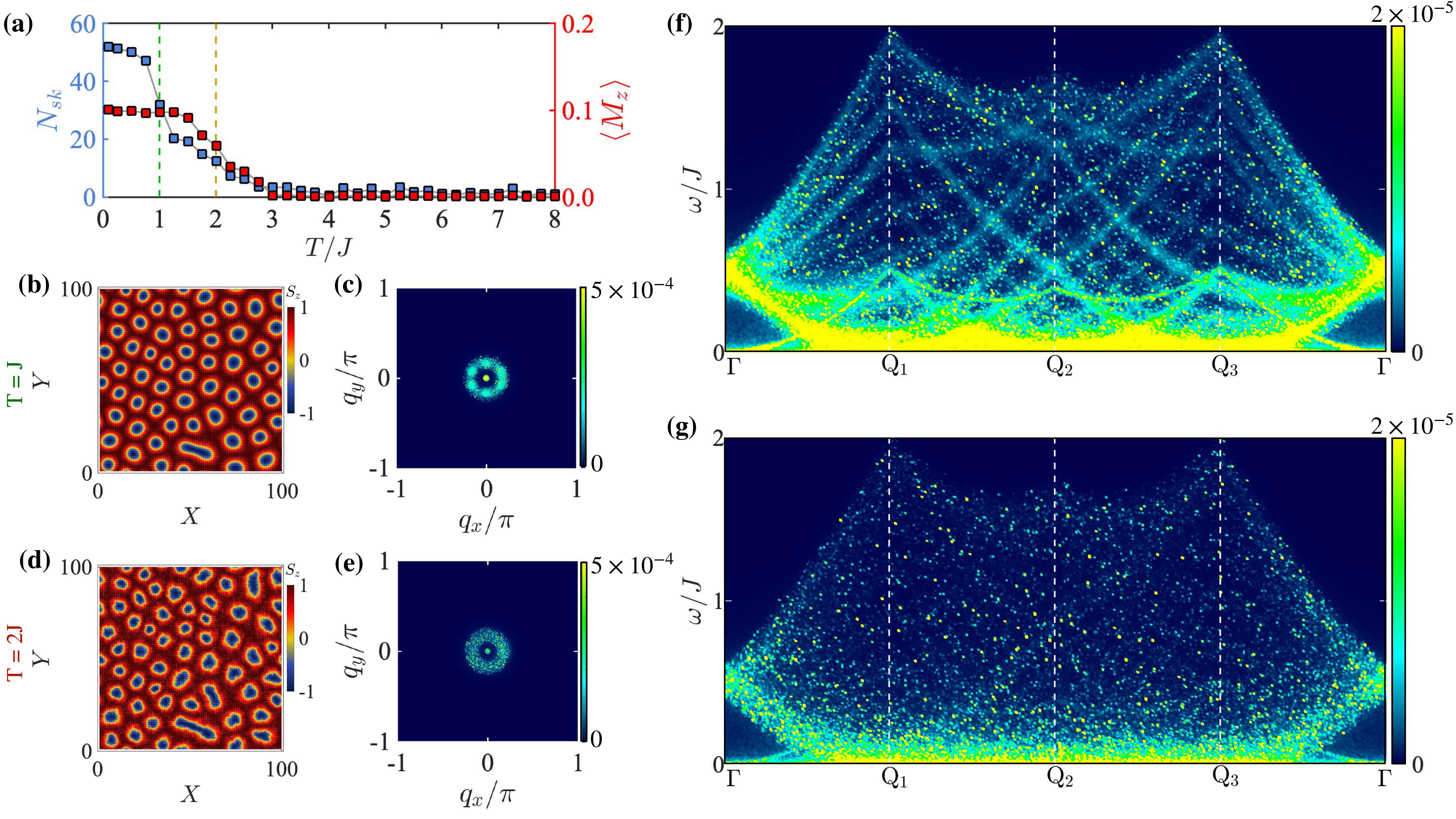,trim=0.0in 0.0in 0.0in 0.0in,clip=false, width=180mm}
\caption{{\bf Spin waves in the diffusive regime for skyrmion crystallization.} (a) Temperature variation of the skyrmion number $N_{sk}$ and the average magnetization component $\langle M_z \rangle$, showing the gradual transition to the triangular SkX below $T\! \lessapprox \!2.5J$. Panels (b) and (c) are the spin configuration and the zero-energy contour of the dynamical spin structure factor $S(\mathbf{q},\omega)$ at temperature $T\!=\!J$ and magnetic field $H_{z}\!=\!0.18J$. Panels (d) and (e) are the spin configuration and the zero-energy contour of $S(\mathbf{q},\omega)$ at temperature $T\!=\!2J$ and magnetic field $H_{z}\!=\!0.18J$.  Panels (f) and (g) show the plots of $S(\mathbf{q},\omega)$ along the momentum path $\Gamma - \mathbf{Q}_1 - \mathbf{Q}_2 - \mathbf{Q}_3 - \Gamma$ at the two temperatures, representatives of the diffusive regime for the SkX crystallization. }
\label{SW_SkXD}
\vspace{-2em}
\end{center}
\end{figure*}

\noindent {\bf Spin waves in the skyrmion crystal.} 
The SkX phase is formed within an intermediate magnetic-field range in which, upon cooling, the skyrmions establish a long-range order to form a triangular crystal. At the DMI strength $D\!=\!0.5J$, the SkX phase is realized within the field range $0.13J\! \lessapprox  \!H_z\! \lessapprox \!0.20J$. We focus on the SkX phase obtained at $H_{z}\!=\!0.18J$, shown in Fig.~\ref{Spin_Config}(b). In Fig.~\ref{SW_SkX}(a)-(f), we show the constant-energy contours of the dynamical spin structure factor $S(\mathbf{q},\omega)$ at different energies. The plot at $\omega\!=\!0$ in Fig.~\ref{SW_SkX}(a) depicts the six-peak structure of the SkX phase, as revealed in elastic neutron-scattering experiments (see \textit{e.g.} Fig.2E in Ref.~\onlinecite{Muhlbauer_Science2009}). The three characteristic momenta of the triangular SkX phase are denoted by $\mathbf{Q}_1$, $\mathbf{Q}_2$ and $\mathbf{Q}_3$. The peak at the $\Gamma$ point originates from the zero-momentum magnetic ordering in the ferromagnetic background of the SkX. The constant-energy contours of $S(\mathbf{q},\omega)$ at different finite energies are shown in Fig.~\ref{SW_SkX}(b)-(f). At a finite energy $\omega \!=\!J$, the seven elastic peaks evolve into circles as shown in Fig.~\ref{SW_SkX}(b). With the increase in energy, these circles grow in size and deviate their shape to rounded squares near energy $\omega \!=\!6J$ (Fig.~\ref{SW_SkX}(d)). At much higher energies, the topology of the energy contour changes (characteristic of the Lifshitz transition), seven pockets appear at the corners of the Brillouin zone (Fig.~\ref{SW_SkX}(e)) and these modes disappear at energies above $\omega \!\approx\!13J$.

In Fig.~\ref{SW_SkX}(g), we present the momentum$-$energy dependence of $S(\mathbf{q},\omega)$ of the SkX phase, the main result of the paper, along the high-symmetry path $\Gamma - \mathbf{Q}_1 - \mathbf{Q}_2 - \mathbf{Q}_3 - \Gamma$. The plot unveils a complex structure for the spin-wave modes, six of which originate from the SkX and another from the ferromagnetic background. The spin-wave modes from the SkX appear to follow a parabolic dispersion relation ($\omega \propto q^2$) in the low-energy regime $\omega \lessapprox 0.2J$, around each elastic peak. As shown in Fig.~\ref{SW_SkX}(h), the spin-wave modes of the SkX along the $\Gamma - {\rm X}$ direction are two-fold degenerate. The rich features shown in Fig.~\ref{SW_SkX}(g) are a remarkable result and its possible experimental confirmation is addressed below.

In the presence of an additional AC magnetic field, three spin-wave modes of the SkX phase were numerically identified in Ref.~[\onlinecite{Mochizuki_PRL2012}]. Two of these three modes are rotational modes (clockwise and counterclockwise) that appear with in-plane AC magnetic field. The third one which was found with out-of-plane AC magnetic field is called the ``breathing'' mode, where the skyrmion core expands and shrinks alternatively. These modes were subsequently reported in experiments using the broadband microwave spectroscopy in the skyrmion-host compounds MnSi, Fe$_{1-x}$Co$_{x}$Si, Cu$_2$OSeO$_3$ and GaV$_4$S$_8$~\cite{Schwarze_NMat2015,Onose_PRL2012,Ehlers_PRB2016}. We speculate that these three modes correspond to the excitations at energy $\omega \! \approx \!0.5J$ at the $\Gamma$ point in  Fig.~\ref{SW_SkX}(g). Previous efforts~\cite{Petrova_PRB2011,Garst_JAPD2017,Santos_PRB2018} analytically obtained the spin-wave modes of the SkX that we explore here. Below, we discuss the feasibility of the detection of the spin-wave modes, obtained in our calculations at constant magnetic field, in inelastic neutron-scattering experiments.\\
\begin{figure}[t]
\begin{center}
\epsfig{file=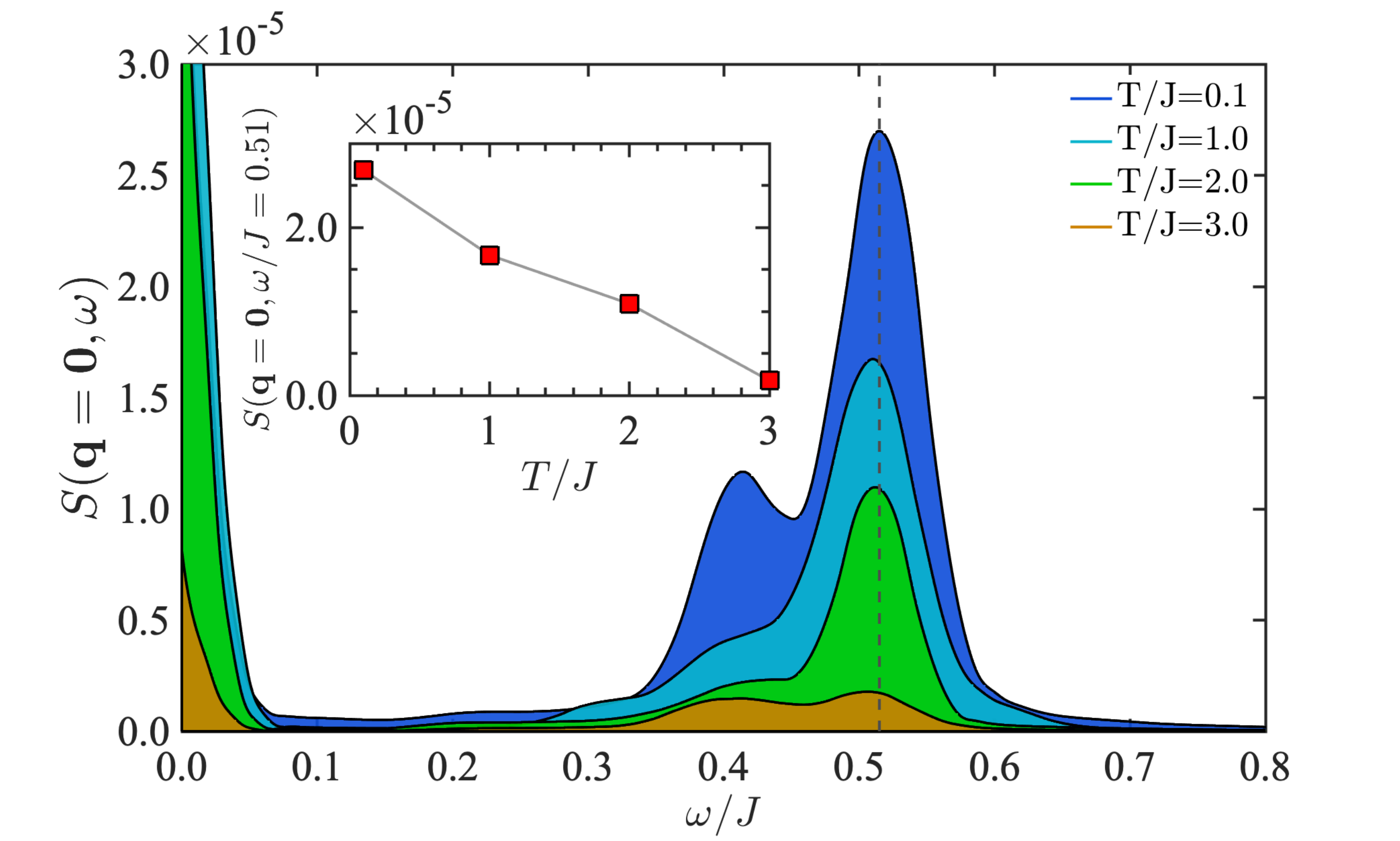,trim=0.4in 0.1in 0.4in 0.1in,clip=true, width=90mm}
\caption{ {\bf Spin-wave mode at zero wave vector across the paramagnet-skyrmion crystal transition.} The energy dependence of the dynamical spin structure factor $S(\mathbf{q\!=\!0},\omega)$ at different temperatures across the phase transition to the skyrmion crystal is shown. The linear decay (inset) of the $S(\mathbf{q\!=\!0},\omega)$ amplitude at energy $\omega/J\!=\!0.51$ suggests the absence of a sharp transition.}
\label{T_Sw}
\vspace{-2em}
\end{center}
\end{figure}

\begin{figure*}[htb!]
\begin{center}
\epsfig{file=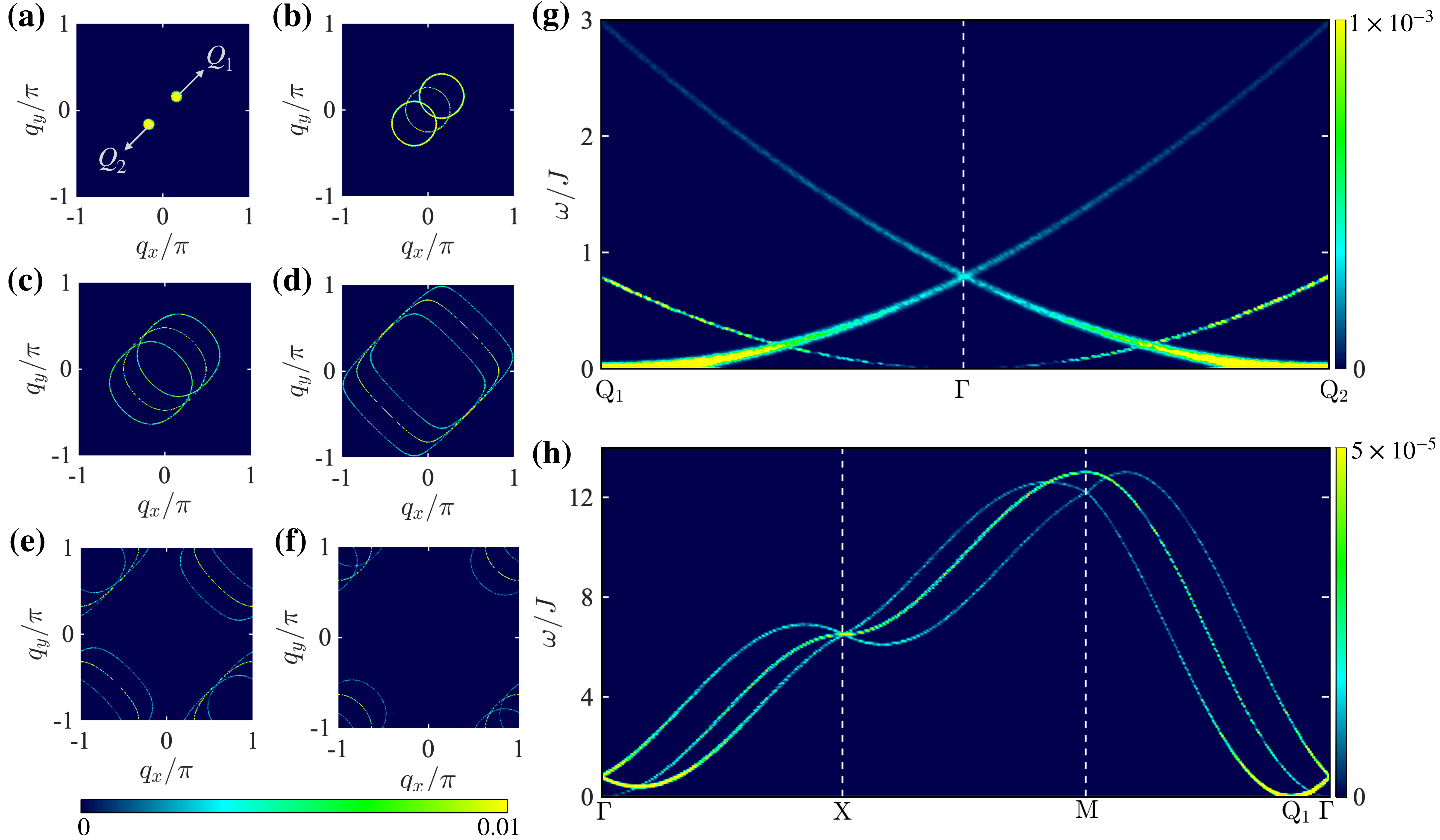,trim=0.0in 0.0in 0.0in 0.0in,clip=false, width=180mm}
\caption{{\bf Spin waves in the spin spiral (SS) phase.} (a)-(f) Constant-energy contours of the dynamical spin structure factor $S(\mathbf{q},\omega)$ calculated in the SS phase, realized at zero magnetic field, at energies (a) $\omega \!=\!0$, (b) $\omega \!=\!J$, (c) $\omega \!=\!3J$, (d) $\omega \!=\!6J$, (e) $\omega \!=\!8J$, and (f) $\omega \!=\!11J$. $\mathbf{Q}_1$ ($=\!-\mathbf{Q}_2$) is the characteristic momentum of the SS phase, as shown in  plot (a). (g)-(h) Plot of $S(\mathbf{q},\omega)$ along the momentum paths (g) $\mathbf{Q}_1 - \Gamma - \mathbf{Q}_2$ and (h) $\Gamma - \rm X - \rm M - \Gamma$. Panel (g) is a symmetrized  expansion of the lower right corner of panel (h).The `soft' spin-wave mode, which arises above a finite energy around the $\Gamma$ point (0,0), is due to the ferromagnetic correlation within the stripes in the SS phase. The two other spin-wave modes arise from the spiral order in the spin configuration.}
\label{SW_SS}
\vspace{-2em}
\end{center}
\end{figure*}

\noindent {\bf Diffusive regime for skyrmion crystallization.} We now focus on the high-temperature regime in which the crystallization of the skyrmions takes place as the temperature is lowered down. In our model, the triangular SkX phase is stabilized below a temperature $T_s\! \lessapprox \! 0.8J$ (which, in Kelvin unit, lies in the range 11~K $\lessapprox$ $T_s$ $\lessapprox$ 35~K, depending on the thickness of the manganite layer), as revealed by the temperature dependence of the skyrmion number $N_{sk}=\frac{1}{4\pi} \int \mathbf{S} \cdot ( \frac{\partial \mathbf{S}}{\partial x} \times \frac{\partial \mathbf{S}}{\partial y} ) dx dy$, plotted in Fig.~\ref{SW_SkXD}(a). The average magnetization component $\langle M_z \rangle$ also becomes nearly constant below $T_s$. However, the development of the skyrmionic correlation begins at temperatures below $T\! \approx \! 2.5J$, much above $T_s$, in the annealing process when the temperature is reduced gradually. The temperature range $0.8J$ $\lessapprox$ $T$ $\lessapprox$ $2.5J$ is the weak crystallization regime at the considered values of $D$ and $H_z$. The snapshots of the spin configuration at two temperatures $T\!=\!J$ and $T\!=\!2J$ are shown, respectively in Fig.~\ref{SW_SkXD}(b) and (d). This diffusive regime may also involve an extended phase of nucleated skyrmions, which we explored in our previous study~\cite{NM_PRB2019}. The constant-energy contour of the dynamical spin structure factor $S(\mathbf{q},\omega)$ at $\omega \!=\! 0$ and at temperature $T\!=\!2J$, in Fig.~\ref{SW_SkXD}(e), shows a ring-shaped profile. At the smaller temperature $T\!=\!J$, in Fig.~\ref{SW_SkXD}(c), the six-peak structure, which is a reminiscent of the triangular crystal of skyrmions, appears with the ring in the background. Such a paramagnetic-to-skyrmion lattice transition has been observed in the transition metal helimagnet MnSi using small-angle neutron scattering, neutron-resonance spin-echo spectroscopy, and microwave spectroscopy experiments~\cite{Kindervater_PRX2019}.

Figures~\ref{SW_SkXD}(f) and (g) show the energy-momentum dependence of $S(\mathbf{q},\omega)$ along the momentum path $\Gamma - \mathbf{Q}_1 - \mathbf{Q}_2 - \mathbf{Q}_3 - \Gamma$ at the two temperatures $T\!=\!J$ and $T\!=\!2J$, respectively. At $T\!=\!J$, the spin-wave modes of the SkX can be resolved well from the merged background of the diffusively-scattered intensities. On the contrary, at $T\!=\!2J$, the ring-shaped elastic profile do not reveal any clear mode of the spin-wave branches. Interestingly, however, the modes near $\omega \! = \!0.51J$ at $\mathbf{q=0}$ persist, despite a smearing in the intensities within a narrow energy range. In Fig.~\ref{T_Sw}, we show the evolution of this $\omega \! = \!0.51J$ mode at different temperatures across the paramagnet-SkX transition and find that the intensity of $S(\mathbf{q},\omega)$ decreases linearly below the temperature $T_s\leq3J$. The absence of a sharp transition in the intensity of $S(\mathbf{q},\omega)$ further indicates the occurrence of a liquid to crystal transition leading to the SkX phase. The measurement of these smeared modes of $S(\mathbf{q},\omega)$ at zero wave vector can be useful to distinguish the elastic neutron-scattering pattern of a SkX from the nearly-similar one of the hexagonal ferrofluids~\cite{Gollwitzer_JPCM2006}, iron oxide nanoparticles~\cite{Fu_Nanoscale2016}, hexagonal magnets~\cite{Takagi_SciAdv2018} and possibly ferromagnetic domains.

The transition from the ring shape to the six-peak structure of elastic $S(\mathbf{q},\omega)$ profile upon reducing temperature in the diffusive regime establishes the fact that fluctuating skyrmionic correlation starts to develop at temperatures much above the SkX-ordering temperature $T_s$ as the precursor of the long-range triangular crystal of skyrmions. Such a precursor phenomenon has been reported in the cubic helimagnets FeGe and MnSi using ac-magnetic susceptibility and magneto‐heat capacity measurements~\cite{Wilhelm_PRL2011,Samatham_PhysStatusSol2013}.\\

\noindent {\bf Spin waves in the spin spiral.} We now turn our attention to the SS phase that appears spontaneously in the absence of any external magnetic field at low temperatures in our model. The SS phase has a single characteristic momentum $\mathbf Q_1$ that is governed by the period of the spiral texture. It is shown in the constant-energy contour of $S(\mathbf{q},\omega)$ at $\omega \!=\!0$ in Fig.~\ref{SW_SS}(a). With an increase in energy, the two elastic peaks at $\mathbf Q_1$ and $\mathbf Q_2=-\mathbf Q_1$, similarly as in the SkX phase, evolve into circles around these characteristic momenta, as shown by the constant-energy contour of $S(\mathbf{q},\omega)$ at $\omega \!=\!J$ in Fig.~\ref{SW_SS}(b). In addition to the two spin-wave modes of the spiral texture, a soft ferromagnetic mode appears at finite energies, as shown by the circle around the $\Gamma$ point in Fig.~\ref{SW_SS}(b). The intensity of this ferromagnetic mode decreases with a decrease in the period of the SS texture, \textit{i.e.} with an increase in the DMI strength $D$.  The constant-energy contours of $S(\mathbf{q},\omega)$ at different energies in Fig.~\ref{SW_SS}(c)-(f) show the spin-wave modes behaving in a similar manner as in the FM or in the SkX phase. 

We show the momentum$-$energy dependence of $S(\mathbf{q},\omega)$ of the SS phase along the momentum path $\mathbf{Q}_1 - \Gamma - \mathbf{Q}_2$ in Fig.~\ref{SW_SS}(g), and along $\Gamma - \rm X - \rm M - \Gamma$ in Fig.~\ref{SW_SS}(h). The plots show two spin-wave modes from the spiral order and one mode from the ferromagnetic order with a bandwidth $\sim$$13J$. The spin-wave modes from the SS phase follow nearly a parabolic dispersion relation ($\omega \propto q^2$) in the low-energy regime $\omega \lessapprox J$. The two-fold degenerate spin-wave excitation at $\omega \! \approx \!0.8J$ at the $\Gamma$ point (Fig.~\ref{SW_SS}(g)) is detectable in microwave spectroscopy experiments.

Theoretical calculations~\cite{Garst_JAPD2017} and inelastic neutron-scattering experiments  in the context of chiral magnets~\cite{Kugler_PRL2015,Schwarze_NMat2015,Grigoriev_PRB2015,Dussaux_NComm2016,Tucker_PRB2016,Portnichenko_NComm2016,Bauer_PRB2017,Turgut_PRB2017,Siegfried_PRB2017,Weiler_PRL2017} analyzed the ``helimagnon'' excitation modes of the SS phase.

A ferromagnetic spin-wave mode, corresponding to the zero wave-vector magnetic order, is expected to robustly emerge from $\omega \!=\!0$ in the case of the conical spin spiral phase, that appears in three-dimensional chiral magnets at zero magnetic field and in two-dimensional oxide interfaces with an in-plane magnetic field~\cite{Garst_JAPD2017}.\\

\begin{figure}[t]
\begin{center}
\epsfig{file=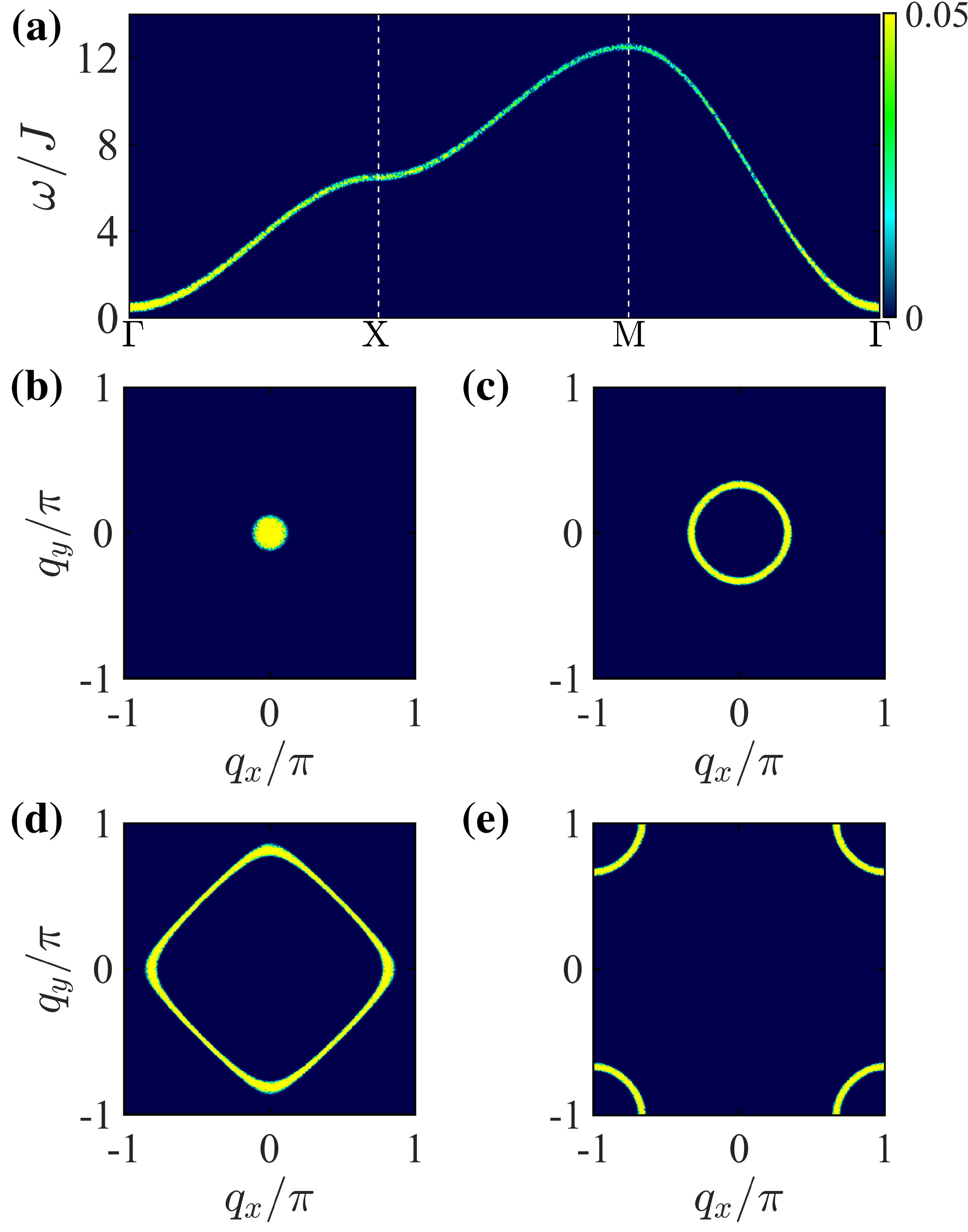,trim=0.0in 0.0in 0.0in 0.0in,clip=false, width=88mm}
\caption{{\bf Spin waves in the field-polarized ferromagnetic (FM) phase.} (a) The dynamical spin structure factor $S(\mathbf{q},\omega)$ calculated in the FM phase, realized at a magnetic field $H_{z}\!=\!0.5J$ ($\approx 10.37$~T), along the momentum path $\Gamma~(0,0) - \rm X~(\pi,0) - \rm M~(\pi,\pi) - \Gamma$. Plots (b) to (e) show the constant-energy contours of $S(\mathbf{q},\omega)$ at energies (b) $\omega \!=\!0.5J$, (c) $\omega \!=\!2J$, (d) $\omega \!=\!6J$, and (e) $\omega \!=\!11J$. }
\label{SW_FM}
\vspace{-1em}
\end{center}
\end{figure}

\noindent {\bf Spin waves in the field-polarized ferromagnet.} Having explored all other magnetic phases in our model, we now discuss, for concreteness, the spin-wave mode of the field-polarized FM phase. We realized this FM phase at a magnetic field $H_z\!=\!0.5J$ in the MC simulations. Figure~\ref{SW_FM}(a) shows the dynamical spin structure factor $S(\mathbf{q},\omega)$ computed along a momentum path $\Gamma~(0,0) - \rm X~(\pi,0) - \rm M~(\pi,\pi) - \Gamma$ in the Brillouin zone. 

The plot describes the well-known spin-wave dispersion relation $\omega \!=\! H_z+4JS(\cos{q_x}+\cos{q_y})$ with a bandwidth $\sim$$12J$ and an energy gap $\Delta \omega \!=\!0.5J$. Figures~\ref{SW_FM}(b)-(e) show the constant-energy contours of the spin-wave dispersion at different energies. The point-like peak in $S(\mathbf{q},\omega)$ at the $\Gamma$ point at energy $\omega \!=\!0.5J$ evolves into a circle with increasing $\omega$ and the circle takes the shape of a rounded square at $\omega \!\approx\! 6J$. With further increase in $\omega$ and four pockets appear at the corners of the Brillouin zone, like in the cases of the SkX and the SS phases. These hole pockets disappear above the upper limit, $\omega \!\approx\!12.5J$, of the energy spectrum. 

The sharpness of the $S(\mathbf{q},\omega)$ dispersion in the frequency$-$momentum space is controlled primarily by the lattice size and the time step. At smaller lattice sizes, the $S(\mathbf{q},\omega)$ peaks appear discontinuously along the dispersion curve. With larger lattice sizes, the increase in the number of peaks form a continuous sharply-defined curve. An analysis of  finite-size effects on the spin-wave dispersion was done in Ref.~\onlinecite{Landau_PRB1994}. On the other hand, a larger time step introduces tails along the $\omega$ axis surrounding the peak positions with an amplitude that decays in an oscillatory manner with distance from the $\omega$ of the peak. A sufficiently small time step reduces these tails significantly from the dispersion curve.  In three-dimensional materials with a finite DMI, the fully-polarized FM phase at a finite magnetic field exhibits an anisotropy in the spin-wave dispersion, the anisotropy being proportional to the DMI strength $D$ and the perpendicular momentum $k_z$~\cite{Garst_JAPD2017,Grigoriev_PRB2015}. In two dimensions, the anisotropy goes away and the spin-wave dispersion is similar to a two-dimensional ferromagnet.\\

\noindent {\bf Discussion\\}
\label{sectionIV}
The presented dynamical spin structure factor of the skyrmion crystal reveals a rich structure of the spin-wave excitations that can be achieved at the maganite/iridate or similar interfaces where a skyrmion crystal is expected to be formed. Its detection in inelastic neutron-scattering experiments can offer an opportunity to effectively probe the topological magnetic textures which is often difficult to deduce conclusively from only the topological Hall effect measurements. The challenge for the successful detection of the SkX spin-wave modes in inelastic scattering experiments lies in obtaining a sufficiently high signal-to-noise ratio from a single interface. For this purpose, multilayer heterostructures, composed of alternate manganite and iridate layers can provide a workable platform. 
We note that with the mass of a typical La$_{1-x}$Sr$_{x}$MnO$_3$/SrIrO$_3$/SrTiO$_3$ superlattice (of thickness 1-10 $\mu$m deposited on a 1$\times$1 cm$^2$ substrate) is $\sim$0.00064 g (with average density $\sim$6.4 g/cm$^3$), implying that a sample mass of 0.1 g can be achieved by stacking $\sim$15-150 of these superlattices.
Considering that LaMnO$_3$ has a molar mass 239.3 g/mol, a 0.1~g
sample contains 0.4 millimoles, while a generic criterion for inelastic neutron scattering experiments is to have
samples of about 6.2 (3.5) millimoles for S=3/2 (S=2) Mn spins. Thus, measurements with these small samples are likely a factor 10 beyond current capabilities, such as the Spallation Neutron Source in the United States, the Institut Laue-Langevin in France, the ISIS neutron source in the United Kingdom, and J-PARC in Japan. There are additional considerations such as neutron absorption, background from the substrate scattering, bandwidth of the excitations, achieving appropriate coverage in reciprocal space, etc for a successful neutron-scattering experiment. However, such experiments are likely  feasible at planned or under construction next generation neutron sources, such as the European Spallation Source or the Second Target Station at the Spallation Neutron Source, Oak Ridge National Laboratory, where improvements in instrumentation and neutron flux are anticipated to result in performance gains in excess of an order of magnitude~\cite{SNS_First_experiments,Instruments_ESS,CSPEC_ESS}.

Although the challenges related to the thin-film geometry of oxide interfaces are considerable with
regards to sample mass, if robust results could be experimentally gathered they will enrich the understanding of the dynamics of the skyrmions which are of high technological interests owing to their potential for wide-spread applications in the high-density data storage~\cite{Everschor_JAP2018} and magnonic devices~\cite{Ma_NanoLett2015}. Another merit of the presented results and the proposed inelastic measurements would be to distinguish the elastic neutron-scattering pattern of a skyrmion crystal from the nearly-similar one of the hexagonal ferrofluids~\cite{Gollwitzer_JPCM2006}, iron oxide nanoparticles~\cite{Fu_Nanoscale2016}, and hexagonal magnets~\cite{Takagi_SciAdv2018}.

To conclude, we theoretically investigated the spin-wave excitations of a two-dimensional Dzyaloshinskii-Moriya magnet, in the context of the two-dimensional interface between lanthanum manganite and strontium iridate, by computing the dynamical spin structure factor. We explore the excitation modes appearing in the SkX, SS, and FM phases obtained using Monte Carlo simulations and Landau-Lifshitz spin dynamics at different magnetic fields. In particular, we focus on the SkX phase which shows a complex interesting structure in the spin dynamical spin structure factor at low temperatures. We also present the spin-wave excitations in the diffusive regime above the SkX-ordering temperature $T_s$ and show that fluctuating skyrmionic correlation develops at temperatures much above $T_s$ as the precursor of the long-range SkX phase, indicative of a liquid-crystal transition. We finally discussed the possible challenges in the experimental detection and propose a multilayer heterostructure geometry that might overcome some of the obstacles to successfully observe these spin-wave modes from the SkX phase at oxide interfaces.\\

\begin{figure}[b]
\begin{center}
\epsfig{file=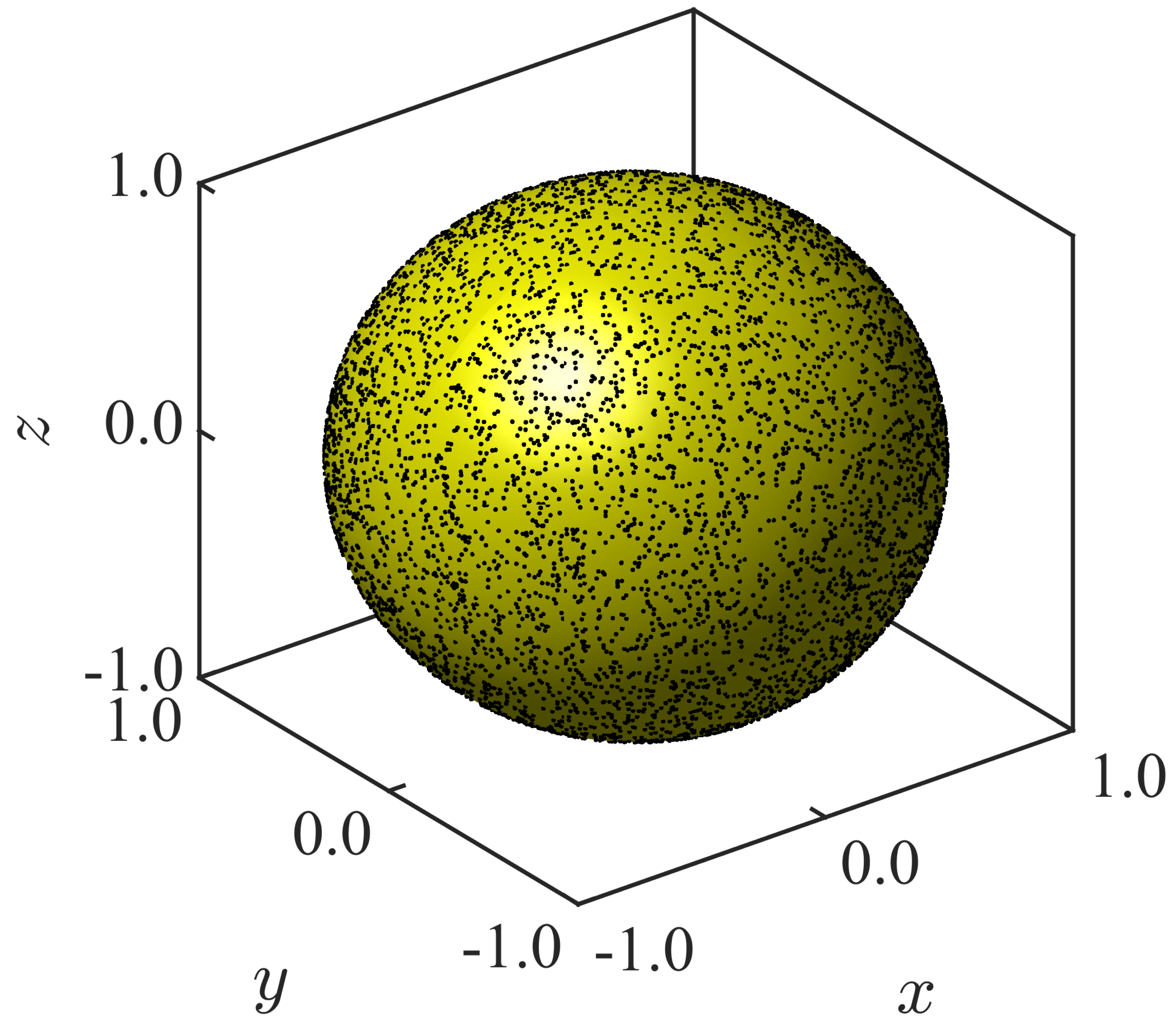,trim=0.0in 0.15in 0.0in 0.12in,clip=true, width=50mm}
\caption{{\bf Monte Carlo sampling in the phase space.} Typical Monte Carlo sampling of $10^{4}$ spin angles ($\theta$,~$\phi$) in the phase space, showing a uniform coverage of the unit sphere by the sampled spin angles (black dots).} 
\label{sampling}
\vspace{-2em}
\end{center}
\end{figure}

\noindent {\bf Methods\\}
{\bf Monte Carlo annealing.} We obtain the spin configurations on a square lattice of size 100$\times$100 with periodic boundary conditions using the Monte Carlo (MC) annealing procedure. For all considered magnetic fields, the annealing process was started at a high temperature $T \!=\! 30J$ with a completely random spin configuration and the temperature was lowered slowly down to a low value $T \!=\! 0.001J$ in $1000$ temperature steps. At each temperature, $10^9$ MC spin updates were performed. In each spin update step, a new spin direction was chosen randomly within a small cone spanned around the initial spin direction. The MC sampling method provides a full and uniform coverage of the phase space spanned by the spin angles, as illustrated in Fig.~\ref{sampling}. The new spin configuration was accepted or rejected according to the standard Metropolis algorithm by comparing the total energies of the previous and the new trial spin configurations, calculated using the Hamiltonian in Eq.~\!(\ref{Hamiltonian}).\\

\noindent {\bf Landau-Lifshitz spin dynamics.} Starting with the  initial conditions obtained from MC simulations at a given temperature, the spins are integrated numerically and averaged over the initial values using the Landau-Lifshitz equations of motion
\begin{align}
\frac{d\mathbf{S_r}(t)}{dt}=\frac{\partial \mathcal{H}}{\partial \mathbf{S_r}(t)} \times \mathbf{S_r}(t),
\label{LL_Eq}
\end{align}
where $\mathcal{H}$ is the Hamiltonian described in Eq.~\!(\ref{Hamiltonian}) and $\mathbf{S_r}(t)$ is the spin vector at position $\mathbf{r}$ and time $t$.
\label{sectionIII}

The central object of investigation is the dynamical spin structure factor $S(\mathbf{q},\omega)$, expressed as 
\begin{align}
S(\mathbf{q},\omega)\!&=\!\sum_{\alpha,\beta}\Big( \delta_{\alpha\beta}-\frac{q_{\alpha}q_{\beta}}{q^2+\eta^2}\Big) \nonumber \\ 
& \times \frac{1}{2\pi N} \sum_{\mathbf{r},\mathbf{r}^{\prime}}e^{-i\mathbf{q}\cdot(\mathbf{r}-\mathbf{r}^{\prime})}
\int_{-\infty}^{\infty}\!dt~e^{-i\omega t} C^{\alpha \beta}(\mathbf{r}-\mathbf{r'},t) 
\label{Sqw}
\end{align}
which is the space-time Fourier transform of the dynamical spin correlation function $C^{\alpha \beta}(\mathbf{r}-\mathbf{r'},t)$
\begin{align}
C^{\alpha \beta}(\mathbf{r}-\mathbf{r'},t)=\Big[\langle S^{\alpha}_{\mathbf{r}}(t)S^{\beta}_{\mathbf{r'}}(0)\rangle \!-\! \langle S^{\alpha}_{\mathbf{r}}(t) \rangle \langle S^{\beta}_{\mathbf{r'}}(0)\rangle \Big],
\end{align}
where $\alpha$ and $\beta$ represent the $x,y,$ and $z$ components. $N$ is the total number of lattice sites and the parameter $\eta \!=\!0.01$ was used to obtain $S(\mathbf{q},\omega)$ at $q\!=\!0$. The factor $( \delta_{\alpha\beta}-\frac{q_{\alpha}q_{\beta}}{q^2+\eta^2})$ was used in Eq.~\!(\ref{Sqw}) to include the off-diagonal components of the  correlation function $C^{\alpha \beta}(\mathbf{r}-\mathbf{r'},t)$ ($\alpha \! \neq \beta$).

While solving Eq.~\!(\ref{LL_Eq}), we express the spins using spherical polar coordinates and evaluate the evolution of the polar and the azimuthal angles $\theta$ and $\phi$. The time integration in Eq.~\!(\ref{LL_Eq}) was carried out using the fourth-order Runge-Kutta method to a maximum time $t_{max}\!=\!1000J^{-1}$ with a time step $\Delta t\!=\!0.001J^{-1}$. The finite time cutoff may introduce small oscillations in $S(\mathbf{q},\omega)$ along the $\omega$ axis, resulting from the time Fourier transform. These oscillations were minimized using the convolution of the spin-spin correlation function with a resolution function in frequency, equivalent to the energy resolution of the SANS spectrometer. A Gaussian broadening function with a broadening parameter $\sigma_{\omega}$ yields the modulated dynamical spin structure factor~\cite{Landau_PRB1994}
\begin{align}
S(\mathbf{q},\omega)\!&=\frac{1}{\sqrt{2\pi} \sigma_{\omega}} \int_{-\infty}^{\infty}S'(\mathbf{q},\omega')exp\Big[-\frac{(\omega-\omega')^2}{2\sigma_{\omega}^2} \Big] d\omega',
\label{Sqw1}
\end{align}
where $S'(\mathbf{q},\omega')$ is the dynamical spin structure factor,  calculated using a time cutoff. In the results presented in this paper, $\sigma_{\omega}\!=\!0.001J$ was sufficient to resolve different spin-wave modes. 

The Gilbert damping may also introduce a broadening of the spin-wave bands. However, the Gilbert damping coefficient ($\alpha$) for manganite systems is very low ( $\sim10^{-4}$) [see, for example, Ref.~\onlinecite{Qin_APL2017}]. These small values of $\alpha$ will introduce only a tiny amount of broadening to the higher-energy spin-wave excitations, and therefore, it can be safely ignored in the context of manganite thin films.

The computed $S(\mathbf{q},\omega)$ was averaged using 50 independent thermalized realizations of the MC spin configurations to reduce the noise arising from small statistical imperfections in those configurations. Increasing this number did not produce noticeable modifications in the results. One advantage of the current approach over linear spin-wave theory techniques is the scope to examine spin dynamics at finite temperatures, where the spin system does not have a well-established long-range order.\\



\noindent {\bf \\Acknowledgments\\}
All members of this collaboration were supported by the U.S. Department of Energy (DOE), Office of Science, Basic Energy Sciences (BES), Materials Sciences and Engineering Division. The authors acknowledge discussion with Elizabeth M. Skoropata and Ho Nyung Lee on the specification of their manganite-iridate thin films.\\



\vspace{-1em}
\def\bibsection{\section*{\refname}} 

%

\end{document}